\documentclass[12pt,a4paper]{article}

\usepackage{latexsym}
\usepackage{graphicx,color}
\usepackage{amsmath}
\usepackage{geometry}
\newcommand\sect[1]{\section{#1}\setcounter{equation}0}

\newcommand\no{\nonumber\\}
\newcommand\eqnb{\begin{eqnarray}}
\newcommand\eqne{\end{eqnarray}}

\usepackage{latexsym}

\begin{document}
{\center{
{\huge On small tension p-branes}\vspace{15mm}\\
Jonas Bj\" ornsson\footnote{jonas.bjornsson@kau.se} and Stephen Hwang\footnote{stephen.hwang@kau.se}\\
Department of Physics\\Karlstad University
\\SE-651 88 Karlstad, Sweden}
\vspace{15mm}\\}

\begin{abstract}
This paper deals with p-branes with small but non-zero tension. 
We prove the existence of canonical transformations, within a perturbation theory, that link specific geometries 
of p-branes to solvable theories, namely string-like and particle-like theories.
The specific shapes correspond to stretched configurations. For configurations linked to string-like
theories one will upon quantization get a critical dimension of (25+p).
\end{abstract}

%%%%%%%%%%%%%%%%%%%%%%%%%%%%%%%%%%%%%%%%%%%%%%%%%%%%%%%%%%%%%%%%%%%%%%%%%%%%%%%%%%%%%%%%%%%

\sect{Introduction and framework}
The theories that describe relativistic $p$-branes are known to be quite complicated. For $p=0,1$ one can solve the equations of 
motion in a flat background, but for $p\geq 2$ one cannot do this. For $p=2$ i.e.\ membranes, one can instead make a reduction 
\cite{Goldstone, Hoppe:1982,deWit:1988ig} which yields a maximally supersymmetric Matrix-theory 
\cite{Flume:1984mn,Claudson:1984th,Baake:1984ie}. 
This theory has many interesting features and is conjectured to give all microscoptic 
degrees of freedome for M-theory \cite{Banks:1996vh,Susskind:1997cw,Sen:1997we,Seiberg:1997ad}. But, for $p\geq 3$ one does not 
know any such reduction.

In this paper we follow another path. We are interested in $p$-branes with small tension and specific 
geometries. These geometries correspond to stretched $p$-branes for which $p-1$, or $p$, are large dimensions. 
For such configurations we will show that one may set up a perturbation theory around a solvable model 
and which makes it possible to solve the equations of motion. The stretched configurations are connected to 
the zero tension limit. For the string, the zero tension limit was first discussed in 
\cite{Schild:1976vq}. Furthermore, such limits have been discussed for D$p$-branes 
as well, \cite{Lindstrom:1997uj,Gustafsson:1998ej,Lindstrom:1999tk}, in which tensile non-interacting strings arise. 
The tensionless limit of $p$- branes are interesting in the same way as the tensionless limit of string theory, being
relevant for a high energy description of the theory.

Here, we will generalize the results of \cite{Bjornsson:2004yp,Bjornsson:2005rv,Bjornsson:2005nh} to hold also 
for $p$-branes. We will show that $p$-branes with small tension and stretched geometries 
can, in general, be described by perturbing free tensile string- or particle-like theories. 
They differ from regular string and particle theories since
the embedding fields depend on $p+1$ world-hypervolume parameters. 

Our main result is to prove that one can,
within perturbation theory, solve exactly the theory by canonically transforming to a free theory. 
A consequence of our result 
is that one can canonically relate a $p$-brane and a $(p-b)$-brane-like theory for arbitrary $b$. These relations 
hold when the tensions are small and the branes are stretched.

Our starting point is the Dirac action \cite{Dirac:1962iy} for the bosonic $p$-brane i flat space-time. This 
theory has $p+1$ constraints due to the reparametrization invariance. The constraints found from this action are
\eqnb
\phi_0
      &=&
          \frac{1}{2}\left[\mathcal P^2 + T_p^2 \det(h_{ab})\right]\approx 0\no
\phi_a
      &=&
          \mathcal P_U \partial_a X^U\approx 0,
\label{p-brane constraints}
\eqne
where $\mathcal{P}_U$ are the canonical momenta, $U=0,\ldots,d-1$ is the spacetime index, $a,b=1,\ldots, p$ are the space-like 
directions of the $p$-brane and $h_{ab}\equiv\partial_a X_U\partial_b X^U$. The constraints are first class and satisfy the
Poisson bracket algebra
\eqnb
\{\phi_a(\xi),\phi_b(\xi')\}
      &=&
          \phi_b(\xi)\partial_a\delta(\xi-\xi')+\phi_a(\xi')\partial_b\delta(\xi-\xi')
      \no
\{\phi_a(\xi),\phi_0(\xi')\}
      &=&
          \left[\phi_0(\xi)+\phi_0(\xi')\right]\partial_a\delta(\xi-\xi')
      \no
\{\phi_0(\xi),\phi_0(\xi')\}
      &=&
          T_p^2p\,\epsilon^{a_1,\ldots,a_p}_{b_1,\ldots,b_p}
          \left[\phi^{b_1}{h_{a_2}}^{b_2}\cdot\ldots\cdot{h_{a_p}}^{b_p}(\xi)
          +(\xi\rightarrow\xi')\right]\partial_{a_1}\delta(\xi-\xi').
\eqne
One could also define a BRST-charge for the theory which yields that it is a rank $p$ theory \cite{Henneaux:1983um}. 

Let us choose a partial gauge by fixing one of the space-like reparametrization invariances. We do this in the same way as in 
\cite{Bjornsson:2004yp} by gauging one of the space-like parameters of the hypervolume to be proportional to one of the 
space-like directions in space-time. Let us for simplicity choose the $p$'th variable and the $d-1$ direction, 
$\chi_1\equiv X^{D-1}-k\xi^p\approx 0$. This will yield the remaining constraints
\eqnb
\phi_0
      &=&
          \frac{1}{2}\left[\mathcal P^2+T_p^2k^2\det(h'_{a'b'})+\frac{1}{k^2}\left(\mathcal P\partial_{p}X\right)^2
          +T_p^2\det(h'_{ab})\right]\approx 0
      \no
\phi_{a'}
      &=&
          \mathcal P_\mu \partial_{a'} X^\mu\approx 0,
\label{constraints(p-1)}
\eqne
where $h'_{ab}\equiv\partial_a X^\mu\partial_bX_\mu$, $a',b'=1,\ldots,p-1$\footnote{We will henceforth skip the prime on 
$h_{ab}$} and where $k$ is a constant. Computing the Poission brackets between the constraints yields that 
they form a closed Poisson bracket algebra. Let us study the algebra and constraints by assuming that $k$ is large and $T_p$ 
small such that $T^2_pk^2= \tilde T_{p-1}^2$ is fixed and finite. We have then that the two first terms of $\phi_0$ and 
$\phi_{a'}$ describe a regular $(p-1)$-brane. Thus, $\tilde T_{p-1}$  can be 
interpreted as the tension for a $(p-1)$-brane-like theory, which has one extra world-hypervolume dependence compared to a regular 
$(p-1)$-brane. The resulting theory is, therefore, a $(p-1)$-brane-like theory with a non-trivial perturbation. 

In order to solve the $(p-1)$-brane theory we need to make further simplifications in order to relate it to a solvable theory. 
We will in the next sections discuss two possibilities, one in which the $p$-brane has a shape such that may 
be related to string-like theory and one in which it may be 
related to a particle-like theory.

%%%%%%%%%%%%%%%%%%%%%%%%%%%%%%%%%%%%%%%%%%%%%%%%%%%%%%%%%%%%%%%%%%%%%%%%%%%%%%%%%%%%%%%%%%%

\sect{A perturbed bosonic string-like theory for stretched $p$-branes}

Let us fix $(p-1)$-constraints by
\eqnb
\chi_{l}
      &=&
          X^{D-l}-k\xi^{p+1-l}\approx 0,
\label{simplegaugefix}
\eqne
where $l=1,\ldots,p-1$. The constraints left are
\eqnb
\phi_0
      &=&
          \frac{1}{2}\left\{\mathcal P^2+T_p^2\sum_{i=0}^{p-1}\left[\binom{p}{i}k^{2i}{h_{a_1}}^{b_1}
          \cdot\ldots\cdot{h_{a_{p-i}}}^{b_{p-i}}\sum_{j_1,\ldots,j_i=2}^{p}
          \epsilon_{b_1,\ldots,b_{p-i},j_1,\ldots,j_i}^{a_1,\ldots,a_{p-i},j_1,\ldots,j_i}\right]
      \right.
      \no
      &+&
      \left.
          \frac{1}{k^2}\sum_{i=2}^{p}\left(\mathcal P_\mu\partial_i X^\mu\right)^2\right\}\approx 0
      \no
\phi_1
      &=&
          \mathcal P_\mu\partial_1X^\mu\approx 0,
\label{stringlikeconstraints}
\eqne
where ${h_{a_i}}^{b_i}=\partial_{a_i}X^\mu\partial^{b_i}X_\mu$ and 
$\epsilon^{a_1,\ldots,a_p}_{b_1,\ldots,b_p}=\frac{1}{p!}\epsilon^{a_1,\ldots,a_p}\epsilon_{b_1,\ldots,b_p}$. 
The constraints satisfy a closed Poisson bracket algebra
\eqnb
\{\phi_1(\xi),\phi_1(\xi')\}
      &=&
          \left[\phi_1(\xi)+\phi_1(\xi')\right]\partial_1\delta(\xi-\xi')\no
\{\phi_1(\xi),\phi_0(\xi')\}
      &=&
          \left[\phi_0(\xi)+\phi_0(\xi')\right]\partial_1\delta(\xi-\xi')
      \no
      &+&
          \frac{1}{k^2}\sum_{i=2}^{p}\phi_1\mathcal P_\mu \partial_i X^\mu(\xi')\partial_i\delta(\xi-\xi')
      \no
\{\phi_0(\xi),\phi_0(\xi')\}
      &=&
          \sum_{i=0}^{p-1}(p-i)\binom{p}{i}T_p^2k^{2i}\sum_{j_1,\ldots,j_i=2}^{p}
          \epsilon^{a_1,\ldots,a_{p-i},j_1,\ldots,j_i}_{1,b_2,\ldots,b_{p-i},j_1,\ldots,j_i}
      \no
      &\times&
          \left[\phi_1{h_{a_2}}^{b_2}\cdot\ldots\cdot{h_{a_{p-i}}}^{b_{p-i}}(\xi)+(\xi\rightarrow\xi')\right]
          \partial_{a_1}\delta(\xi-\xi')
      \no
      &+&
          \sum_{i=2}^p\frac{2}{k^2}\left[\left(\mathcal P \partial_i X\right)\phi_0(\xi)+(\xi\rightarrow\xi')\right]
          \partial_i\delta(\xi-\xi').
\label{stringlikealgebra}
\eqne
Eqs.\ (\ref{stringlikeconstraints}) and (\ref{stringlikealgebra}) are exact expressions for the $p$-brane i.e.\ they  
hold without any assumptions beeing made. The two remaining contraints are first-class, corresponding to the two remaining 
reparametrizations, of which one is time-like. We will in the rest of the section assume $k$ to be large and $T_p$ small 
such that $T_p^2k^{2(p-1)}\equiv\tilde T_1^2$ is fixed, even in the limit $k\rightarrow \infty$ and $T_p\rightarrow 0$. 
Then $\tilde T_1$ can be interpreted as the tension of a string-like theory 
with trivial dependence on the other world-hypervolume parameters. The constraints are of the form 
$\phi_0 = \frac{1}{2}\left[\mathcal{P}^2+\tilde T_1^2\left(\partial_1 X\right)^2\right]+g\left(\ldots\right)$ and 
$\phi_1 = \mathcal{P}_\mu\partial_1X^\mu$ where $g\ll 1$, so that the additional dependence on the world-hypervolume
can be treated perturbatively. As $X^{D-l}=k\xi^{p+1-l}$, we see that $X^{D-l}$ will be large, so that the perturbation
theory assumes brane-shapes with $p-1$ large dimensions.
We will proceed as in \cite{Bjornsson:2005rv} solving the theory 
by successive canonical transformations. First, we gauge fix completely by
\eqnb
\chi_0 
      &=&
          \mathcal P^+ - 1 \approx 0
      \no
\chi_1
      &=&
          X^+ - \xi^0      \approx 0,
\label{gaugefix.string}
\eqne
where we have defined lightcone coordinates by $A^{\pm}\equiv \frac{1}{\sqrt{2}}\left(A^1\pm A^0\right)$. Furthermore, we have set 
$\tilde T_1=1$. The Hamiltonian we take as the momentum in the $\mathcal P^-$ direction,
\eqnb
H
      &=&
          -\int d^p\xi\mathcal P^-.
\eqne
To determine $\mathcal P^-$ one uses eq.\ (\ref{gaugefix.string}) in eq.\ (\ref{stringlikeconstraints}). One then follows the 
steps in \cite{Bjornsson:2005rv}, solving the unperturbed theory, i.e.\ the string-like theory. The unperturbed Hamiltonian 
is of the form
\eqnb
H_0
      &=&
          \frac{1}{2}\sum_{(a),m}\left(\alpha^{(-a)}_{-m}\alpha^{(a)}_m+\tilde\alpha^{(-a)}_{-m}\tilde\alpha^{(a)}_m\right),
\eqne
where $(a)=(I;n_i)$ and $(-a)=(I;-n_i)$. $n_i$ comes from the the dependence on the additional world-hypervolume parameters. 
As the unperturbed Hamiltonian is of the same form as in \cite{Bjornsson:2005rv}, one can use the results in the 
paper to show that to any order in perturbation theory there exists a canonical transformation which maps the 
perturbed Hamiltonian to the unperturbed one. As an example, a generic term to any finite order has the form
\eqnb
H_N
      &=&
          \sum_{r}\sum_{j=0}^{M} q^{(-a_1)}\cdot\ldots\cdots q^{(-a_j)}H_{(r)}^{(a_1),\ldots,(a_j)},
\eqne
where $H_{(r)}$ has modenumber $r$ in the $\xi^1$-direction. The part of the infinitesimal canonical transformation which solves the 
equation
\eqnb
\{H_0,G_N\}
      &=&
          -H_N,
\eqne
needed to transform away $H_N$ is
\eqnb
G_N
      &=&
          \sum_{r\neq 0}\sum_{j=0}^M\sum_{k=0}^{j}\left\{k!\binom{j}{k}\left(2\alpha_0^{(a_1)}\right)
          \cdot\ldots\cdot\left(2\alpha_0^{(a_k)}\right)q^{(a_{k+1})}\cdot\ldots\cdot q^{(a_{j})}
      \right.
      \no
      &\times&
      \left.
          H_{(r)}^{(a_1),\ldots,(a_j)}\left(\frac{i}{r}\right)^{k+1}\right\}+\sum_{j=0}^M\sum_{k=0}^{j}\left\{\binom{j}{k}K^k
          \left(2\alpha_0^{(a_1)}\right)\cdot\ldots\cdot\left(2\alpha_0^{(a_k)}\right)
      \right.
      \no
      &\times&
      \left.
          q^{(a_{k+1})}\cdot\ldots\cdot q^{(a_{j})}H_{(0)}^{(a_1),\ldots,(a_j)}\right\},
\eqne
where $K\equiv k_Iq^I/\left(2k_I\alpha^I_0\right)$ and $k_I$ is some fixed vector such that $k_I \alpha^I_0\neq 0$. This 
shows, by construction, that one can find, to all finite orders, canonical transformations  that
map the perturbed Hamiltonian to the unperturbed string-like one. 
We could here also have taken another path by defining a BRST charge for the constraints in 
eq.\ (\ref{stringlikeconstraints}). By using results on the BRST cohomology \cite{Bjornsson:2005nh},
one can again prove the canonical equivalence. The end result is the same but not 
as explicit as above.

%%%%%%%%%%%%%%%%%%%%%%%%%%%%%%%%%%%%%%%%%%%%%%%%%%%%%%%%%%%%%%%%%%%%%%%%%%%%%%%%%%%%%%%%%%%

\sect{Particle-like theories from stretched $p$-branes}

In this section we fix all but one of the constraints and show how one connect the theory to a particle-like theory. Fix the 
gauge as in eq.\ (\ref{simplegaugefix}), but now with $l=1,\ldots,p$. The only constraint left is the Hamiltonian constraint 
which generates time-like residual reparametrizations
\eqnb
\phi_0
      &=&
          \frac{1}{2}\left\{\mathcal P^2+T_p^2\sum_{i=0}^{p}\left[\binom{p}{i}k^{2i}{h_{a_1}}^{b_1}\cdot\ldots\cdot
          {h_{a_{p-i}}}^{b_{p-i}}
      \right.\right.
      \no
      &\times&
      \left.\left.
          \sum_{j_1,\ldots,j_i=1}^{p}\epsilon_{b_1,\ldots,b_{p-i},j_1,\ldots,j_i}^{a_1,\ldots,a_{p-i},j_1,\ldots,j_i}\right]
          +\frac{1}{k^2}\sum_{i=2}^{p}\left(\mathcal P_\mu\partial_i X^\mu\right)^2
      \right\}\approx 0
\label{constraint.particle}
\eqne
and satisfies the algebra
\eqnb
\{\phi_0(\xi),\phi_0(\xi')\}
      &=&
          \frac{2}{k^2}\sum_{i=1}^{p}\left[\phi_0\mathcal P_\mu \partial_i X^\mu(\xi)
          +\phi_0\mathcal P_\mu \partial_i X^\mu(\xi')\right]\partial_i\delta(\xi-\xi').
\eqne

If one chooses $T_p\ll 1$ and $k\gg 1$ such that $T^2_pk^{2p}=m^2$ is fixed and non-zero, the constraint is of the form 
$\phi_0=\frac{1}{2}\left(\mathcal{P}^2+m^2\right)+g\left(\ldots\right)$ where $g\equiv k^{-2}\ll 1$. 
Therefore, the unperturbed theory describes an infinite 
set of non-interacting particles. By eq.~(\ref{simplegaugefix}) this requires $p$ large dimensions of the brane.

We will now show that one can map the perturbed theory to the unperturbed one by canonical 
transformations in the same manner as in the previous section. We fix, therefore, 
the gauge completely by
\eqnb
\chi_0
      &=&
          X^0-\xi^0\approx 0.
\eqne
The Hamiltonian for the theory can be choosen to be proportional to $\mathcal P^0$
\eqnb
H
      &=&
          \int d^p\xi \mathcal P^0
      \no
      &=&
          \int d^p\xi \sqrt{\mathcal P^2+m^2+gA}
\label{hamiltonian.gaugefixparticle}
\eqne
where
\eqnb
A
      &=&
          m^2\sum_{i=1}^{p}\left[g^{i-1}\binom{p}{i}{h_{a_1}}^{b_1}\cdot\ldots\cdot{h_{a_{p-i}}}^{b_{p-i}}
          \sum_{j_1,\ldots,j_i=1}^{p}\epsilon_{b_1,\ldots,b_{p-i},j_1,\ldots,j_i}^{a_1,\ldots,a_{p-i},j_1,\ldots,j_i}\right]
          +\sum_{i=1}^{p}\left(\mathcal P_\mu\partial_i X^\mu\right)^2.
\eqne
Since $k\gg 1$ we have $g\ll 1$, one can expand eq.\ (\ref{hamiltonian.gaugefixparticle}) to get
\eqnb
H
      &=&
          \int d^p\xi \sqrt{\mathcal P^2+m^2}\sum_{j=0}^{\infty}\binom{1/2}{j}g^j\left(\frac{A}{\sqrt{\mathcal P^2+m^2}}\right)^j
\eqne
where the unperturbed part of the Hamiltonian is $H_0=\sqrt{\mathcal P^2+m^2}$. $H_0$ satisfies
\eqnb
\left\{H_0,\mathcal{P}^I\right\}
      &=&
          0
      \no
\left\{H_0,X^I\right\}
      &=&
          -\mathcal{P}^I\frac{1}{\sqrt{\mathcal{P}^2+m^2}},
\eqne
We can define $K_2\equiv\frac{1}{k_I\mathcal{P}_0^I}\int d^p\xi\left(k_IX^I\right)\sqrt{\mathcal{P}^2+m^2}$. It
satisfies $\left\{H_0,K_2\right\}=-1$, provided $k_I\mathcal{P}_0^I \neq 0$,  which we assume. We now have 
all the tools needed to determine the solution to all finite orders in perturbation theory. The perturbation will, to all 
orders, be polynomials of $X^I$ and derivatives of $X^I$. Furthermore, it will involve polynomials which have zero Poisson 
bracket with the Hamiltonian, $\mathcal P^I$, derivatives of $\mathcal P^I$, $1/\sqrt{\mathcal P^2+m^2}$ and 
$\left(k_I P_0^I\right)^{-1}$. To simplify the problem, one can make a Fourier expansion of the variable dependence of the 
fields. Let us show the explicit solution to a generic term
\eqnb
H_N
      &=&
          \sum_{j=0}^M X_{(-a_1)}\cdot\ldots\cdot X_{(-a_j)}H_N^{(a_1),\ldots,(a_j)}
\eqne
where the index $(a)$ is a collective index for $(I;n_i)$ and $(-a)=(I;-n_i)$. This term may be transformed away by a canonical 
transformation generated by
\eqnb
G_N
      &=&
          \sum_{j=0}^M\sum_{k=0}^j \frac{1}{k+1}\binom{j}{k}K_2^{k+1}
          \left(\int d^p\xi\frac{\mathcal P_{I_1}}{\sqrt{\mathcal P^2+m^2}}\exp\left(in_{1,\,l}\xi^l\right)\right)\cdot\ldots
      \no
      &\times&
          \left(\int d^p\xi\frac{\mathcal P_{I_k}}{\sqrt{\mathcal P^2+m^2}}\exp\left(in_{k,\,l}\xi^l\right)\right)
          X_{(-a_{k+1})}\cdot\ldots\cdot X_{(-a_j)}H_1^{(a_1),\ldots,(a_j)}.
\eqne
We have thus shown that the perturbed Hamiltonian can be mapped to the unperturbed one by successive canonical transformations. 

%%%%%%%%%%%%%%%%%%%%%%%%%%%%%%%%%%%%%%%%%%%%%%%%%%%%%%%%%%%%%%%%%%%%%%%%%%%%%%%%%%%%%%%%%%%

\sect{Further results and quantization}

In the previous sections we have shown that stretched $p$-branes are 
canonically equivalent, within a perturbation theory, to either a free string-like theory, or to a free particle-like theory.
We can use this result to show the canonical equivalence between a stretched $p$-brane with small tension and a stretched
$(p-b)$-brane-like theory with a small tension.

The stretched $(p-b)$-brane is canonically equivalent to 
an unperturbed string- or partice-like theory for small $(p-b)$-brane tension. This requires
$p-b-1$, or $p-b$ for the particle case,  large dimensions. This result holds clearly also if we 
add trivial dependence on additional world-hypervolume parameters. Thus, we can canonically link a
$(p-b)$-brane-like theory to a string- or particle-like theory.
As the stretched $p$-brane is canonically equivalent to the 
string- or particle-like theory for small tensions as well,
one can use the inverse canonical transformation to show that the $p$-brane and the $(p-b)$-brane-like theories 
are canonically equivalent. This requires $p-1$, or $p$, large dimesions for the $p$-brane and 
$p-b-1$, or $p-b$, large dimensions for the $(p-b)$-brane-like theory.

Let us also briefly discuss the quantization of the p-branes and, furthermore, consider the case where the unperturbed theory is 
the string-like theory. The quantization procedure follows, straightforwardly, from \cite{Bjornsson:2005rv}. One defines 
a vacuum and a normal ordering from the solutions of the free string-like theory. Then one makes the inverse infinitesimal 
canonical transformation, which classically is equal to the $p$-brane theory up to some order, and which defines a quantum theory for 
the $p$-brane perturbatively. This procedure yields, among other things, a non-trivial ordering of the operators. This ordering
will imply the existence of a critical dimension coming from the one for the bosonic string. One finds the 
critical dimension for a consisitent quantum theory of the $p$-brane to be $d=25+p$. A further result, is the mass-spectrum, 
which as shown in \cite{Bjornsson:2005rv}, will get a constant shift compared to the string spectrum. This holds to all non-zero 
orders in perturbation theory. 

For the case when the unperturbed theory is particle-like, we will not get a critical dimension. 
Thus, the two possibilities that we have treated are not equivalent at
the quantum level. This is perhaps not surprizing as we quantize around two different types of geometries.
\vspace{0.8cm}

\noindent
{\bf{Acknowledgements.}}
\\
\noindent
S.H. is partially supported by the Swedish Research Council under project no.\ 621-2005-3424.

\end{document}